\documentclass[pre,twocolumn,a4paper,showpacs]{revtex4-1}
\usepackage{amssymb}
\usepackage{wasysym}
\usepackage{graphicx}
\usepackage{epsfig}
\usepackage[caption=false]{subfig}
\begin{document}

\title{The collapse of  ecosystem engineer populations}

\author{Jos\'e F. Fontanari}

\affiliation{Instituto de F\'{\i}sica de S\~ao Carlos,
  Universidade de S\~ao Paulo,
  Caixa Postal 369, 13560-970 S\~ao Carlos, S\~ao Paulo, Brazil}

\begin{abstract}

Humans are the ultimate ecosystem engineers  who have  profoundly transformed the world's landscapes in order to
 enhance their survival. Somewhat paradoxically,  however, sometimes the unforeseen effect
of this ecosystem engineering is the very collapse of the  population it intended to protect. Here we use a spatial version of a
standard population dynamics model of ecosystem engineers to study the   colonization of unexplored virgin territories by
a small settlement of engineers. We find that   during the expansion phase  the population density reaches values much higher than those
the environment can support in the equilibrium situation. When the colonization front  reaches the boundary of the available space, the population density plunges sharply and attains its equilibrium value. The collapse takes place without warning  and happens just after the population reaches its peak number. We conclude that overpopulation  and the consequent collapse of an expanding population of ecosystem engineers is a natural consequence of the nonlinear feedback between the population and environment variables.
\end{abstract}

\maketitle

\section{Introduction}

There is hardly a landscape on Earth  that has not been modified by past living beings as a result of  the natural feedback between organisms and  environment, whose study  was initiated  by Darwin in his last scientific  book \cite{Darwin_81}. A  recent alternative viewpoint - niche construction or ecosystem engineering - acknowledges a  more active role of some species, the so-called ecosystem engineers,  in modifying 
their environments to enhance their survival  \cite{Odling_03}. For instance, beavers  -- an oft-mentioned example of  ecosystem engineer --  cut  trees, build  dams, and  create ponds, thus giving  rise to  new and safe landscapes  \cite{Wright_04}. In fact,
 since  the  areas flooded by  dams  increase the distance beavers  can travel by water, which is safer than traveling by land, 
 the modified landscape results in a net increase of the beavers' survival expectations   \cite{Dawkins_82}. Whereas the issue  whether beavers and other nonhuman species qualify as ecosystem engineers is disputable \cite{Scott_13}, nobody contends that humans are the paramount ecosystem engineers, who have  shaped
the world  into an (arguably) more hospitable place for themselves,  most often with unlooked-for effects \cite{Smith_07}.  An  extreme
 unforeseen effect of the engineering of landscapes, which is nonetheless ubiquitous in the history of civilizations, is the collapse of  human societies caused by habitat destruction and overpopulation, among other factors \cite{Diamond_05}.

 The feature of the population dynamics of ecosystem engineers that makes it well suited to model human populations inhabiting isolated
 areas (e.g., islands and archipelagos) is that the growth of the  population is determined by the availability of usable habitats, which in turn are created by the engineers through the modification, and consequent  destruction, of virgin habitats. From a mathematical perspective, this feedback loop results in a density-dependent carrying capacity.  This feature is the  core of the continuous-time, space-independent  mathematical model that Gurney and Lawton proposed to describe the population dynamics of ecosystem engineers  \cite{Gurney_96}. The Gurney and Lawton model   considers the quality of the habitats as dynamic variables,   in addition to the density of engineers. There are three different types of habitats: virgin, usable (or modified) and degraded habitats. The transition from the virgin to the usable habitat is effected only  in the presence of engineers.  The modified habitats then degrade and eventually recover to become virgin habitats again.  Virgin and degraded habitats are unsuitable for the growth of the engineer population. This ecosystem engineering approach  seems way more suitable to study the interplay between humans and their environment than the traditional predator-prey framework used in previous studies \cite{Brander_98,Motesharrei_14}, although both approaches exhibit the characteristic population cycles that reflect the opposite interests of the  interacting  parts  \cite{Turchin_03a}.
 
The  Gurney and Lawton model becomes more effective (and instructive) to simulate the human-environment interaction if we use its recently proposed  spatial formulation,  where an initial small settlement of engineers is surrounded by vast areas (patches) of virgin habitats, and a fraction of the engineers are allowed to move between neighboring patches \cite{Franco_17}. In time, a patch  is an   ecosystem, say, an island, that can potentially exhibit  all three types of habitats as well as the  engineer population, simultaneously.  Hence
the group of patches can be thought of as an archipelago.  In this contribution we focus on the characterization of the speed of the  colonization front and on global demographic quantities such as the  total mean density of engineers.

Our main finding is that  overpopulation is a natural outcome of the population dynamics of ecosystem engineers during the expansion phase to colonize the unexplored virgin patches. When all patches are explored, the population density plunges sharply towards its (local) equilibrium value. The collapse takes place just after the population reaches its peak number.  This surprising outcome, which results from the nonlinear feedback between engineers and environment,   could hardly be predicted without  mathematics, thus lending credence to the tenets  of the discipline Cliodynamics that advocates  the mathematical modeling of historical processes  \cite{Turchin_03b,Turchin_08}. 
 
The rest of the paper is organized as follows. In Section \ref{sec:model} we offer an overview of the discrete  time version of Gurney and Lawton model of ecosystem engineers. In particular,  we present the  recursion equations  that govern  the local (single-patch) dynamics and summarize the relevant findings regarding the stability of the fixed-point solutions \cite{Franco_17}.   The coupled map lattice  version of  the  discrete  time  model is then  introduced  in Section  \ref{sec:spatial}.  The numerical solution of the coupled map  lattice equations  is presented and discussed in Section  \ref{sec:res} for the case the patches  are arranged in a chain with reflective boundary conditions and the model parameters are set such that the local dynamics is attracted by a nontrivial  fixed point.  The focus is on the colonization scenario where
an initial settlement of engineers placed in the central patch of the chain is allowed to disperse to neighboring patches. 
Finally, Section \ref{sec:disc} is reserved to our concluding remarks.

\section{The discrete time version of the Gurney and Lawton model }\label{sec:model}

As pointed out, Gurney and Lawton have modeled the local population dynamics of ecosystem engineering using a continuous-time model \cite{Gurney_96}. Here we present a brief overview of a discrete time version of that model \cite{Franco_17} that can be easily extended to incorporate the spatial dependence of the engineer population as well as of the habitat variables, following the seminal works on insect and host-parasitoid systems \cite{Hassell_91,Comins_92} (see \cite{Rodrigues_11,Mistro_12,Rodrigues_15} for more recent contributions).
 
We begin by  assuming that the population of engineers at generation $t$ is composed of $E_t$ individuals and that each engineer requires a unit of usable habitat to survive. Denoting by $H_t$ the  number of units of usable habitats available at generation $t$, so that in the    equilibrium  regime one has $\lim_{t \to \infty} E_t/H_t =1$,  we can use Ricker model \cite{Murray_03}  to write the expected number of engineers at generation $t+1$ as  

\begin{equation}\label{EE1}
E_{t+1} = E_t \exp \left [ r \left ( 1 - E_t/H_t \right ) \right ],
\end{equation}
 
\noindent where  $r$ is the intrinsic growth rate  of  the population of engineers  and $H_t$ plays the role of a time-dependent carrying capacity for the population of engineers.

The essential ingredient of  the Gurney and Lawton model, which sets it apart from the other population dynamics models \cite{Turchin_03}, is the requirement that usable habitats be created by engineers working on virgin habitats. In particular, if we assume that there are  $V_t$ units of virgin habitats at generation $t$, then the fraction $C \left ( E_t \right ) V_t$ of them will be  transformed in usable habitats at the next generation, $t+1$. Here  $C \left ( E_t \right )$ is any function that satisfies $ 0 \leq C \left ( E_t \right )  \leq 1$ for all $E_t$ and $C \left ( 0 \right ) = 0$. Clearly, this function measures the efficiency of the engineer population to build usable habitats from the raw materials provided by the  virgin habitats.  

Usable habitats   decay into degraded habitats that are useless to the engineers, in the sense they lack the raw materials needed to build usable habitats. Let  $\delta H_t$ denote the fraction of usable habitats that
decay to degraded habitats in one generation, where $ \delta \in \left [ 0, 1 \right ]$ is the decay probability.  Then the expected number of units of usable habitats at generation $t+1$ is simply

\begin{equation}\label{HH1}
H_{t+1} = \left ( 1 - \delta \right ) H_t + C \left ( E_t \right ) V_t.
\end{equation}

\noindent   At first sight, one might think that the decay probability $\delta$  should be  density dependent (i.e., $\delta = \delta \left ( E_t \right ) $), particularly in the case  the habitat degradation resulted from the overexploitation of resources. However, in the  Gurney and Lawton model the resources are represented by the  virgin habitats, whose  probability of change into usable habitats   is in fact density dependent, $C = C \left ( E_t \right )$.  For example, in an island scenario, the virgin habitats can be thought of as the native forests whereas the usable habitats are  the lands cleared for crops,   whose degradation, due mainly to erosion and soil depletion of nutrients, is more suitably modeled by  
 a constant decay  probability $\delta$, rather than by a density-dependent one. 

Degraded habitats will eventually recover and become virgin habitats again.   Denoting the fraction of degraded habitats that recover to virgin habitats  in one generation by $\rho D_t$  we can write 

\begin{equation}\label{DD1}
D_{t+1} = \left ( 1 - \rho \right ) D_t + \delta H_t,
\end{equation}
 
\noindent  where $ \rho \in \left [ 0, 1 \right ]$ is the recovery probability. Finally, the recursion equation for the expected number of units of virgin habitats is simply

\begin{equation}\label{VV1}
V_{t+1} =    \left [ 1 - C \left ( E_t \right ) \right ]  V_t +  \rho  D_t .
\end{equation}
 
As expected,  $ V_{t+1} + H_{t+1}  + D_{t+1} = V_{t} + H_{t}  + D_{t}   = T$, where $T$ is  the (fixed) total store of habitats (e.g., the
area of the island).  Hence we can define the habitat fractions $v_t \equiv V_t/T$, $h_t \equiv H_t/T$ and $d_t \equiv D_t/T$ that satisfy $v_t + h_t + d_t =1$ for all $t$. In addition, we define the density of engineers $e_t = E_t/T$ which, differently from the habitat fractions, may take on values greater than 1. In terms of these  intensive quantities, the above recursion equations are rewritten as

\begin{eqnarray}
e_{t+1} &  =  & e_t \exp \left [ r \left ( 1 - e_t/h_t \right ) \right ]  \label{e} \\
h_{t+1} & = & \left ( 1 - \delta \right ) h_t + c \left ( e_t \right ) v_t  \label{h} \\
v_{t+1} & = &  \rho \left ( 1 - v_t -h_t  \right ) + \left [ 1- c \left ( e_t \right ) \right ] v_t, \label{v}
\end{eqnarray}

\noindent  where we have used $ d_t = 1 - v_t - h_t$ and $c \left ( e_t \right ) \equiv C \left ( T e_t  \right )$.

To complete the model we must specify  the density-dependent probability  $c \left ( e_t \right )$, which measures the engineers' efficiency to transform the virgin habitats into usable ones. The function $c \left ( e_t \right )$ incorporates  the collaboration and communication strategies   that allowed the engineer ecosystems to build  collective structures (e.g., termite mounds and  anthills), which  are their  solutions to the external and internal threats to their survival \cite{Fontanari_14,Francisco_16,Reia_17}. For humans,  this function  incorporates the beneficial effects (from their  perspective) of the technological advancements  \cite{Basalla_89} that allowed a more efficient harvesting of natural resources. Alternatively,  $c \left ( e_t \right )$ can be viewed as a density-dependent resource depletion probability. Here we  consider the function

\begin{equation}\label{c}
 c \left ( e_t \right ) = 1 - \exp \left ( - \alpha e_t \right ),
\end{equation}

\noindent  where $\alpha > 0$ is the productivity  parameter, which measures the efficiency of the engineers in transforming  natural resources into useful goods. 
For  $\alpha \ll 1$ we have  $c \left ( e_t \right ) \approx  \alpha e_t$ and  the term responsible  for the  depletion of virgin habitats in eq.\ (\ref{v})  becomes  $ \alpha e_t v_t$, indicating a low-technological organization where, in a finite population scenario, it would be necessary  the direct contact between one engineer and one unit of virgin  habitat in order to  transform it in one unit of usable habitat \cite{Gillespie_76}.  For $\alpha \gg 1$, however, a few  engineers can transform all the available virgin
habitats  in just  a single generation.

\begin{figure}[h]
\centering
\includegraphics[width=0.9\linewidth]{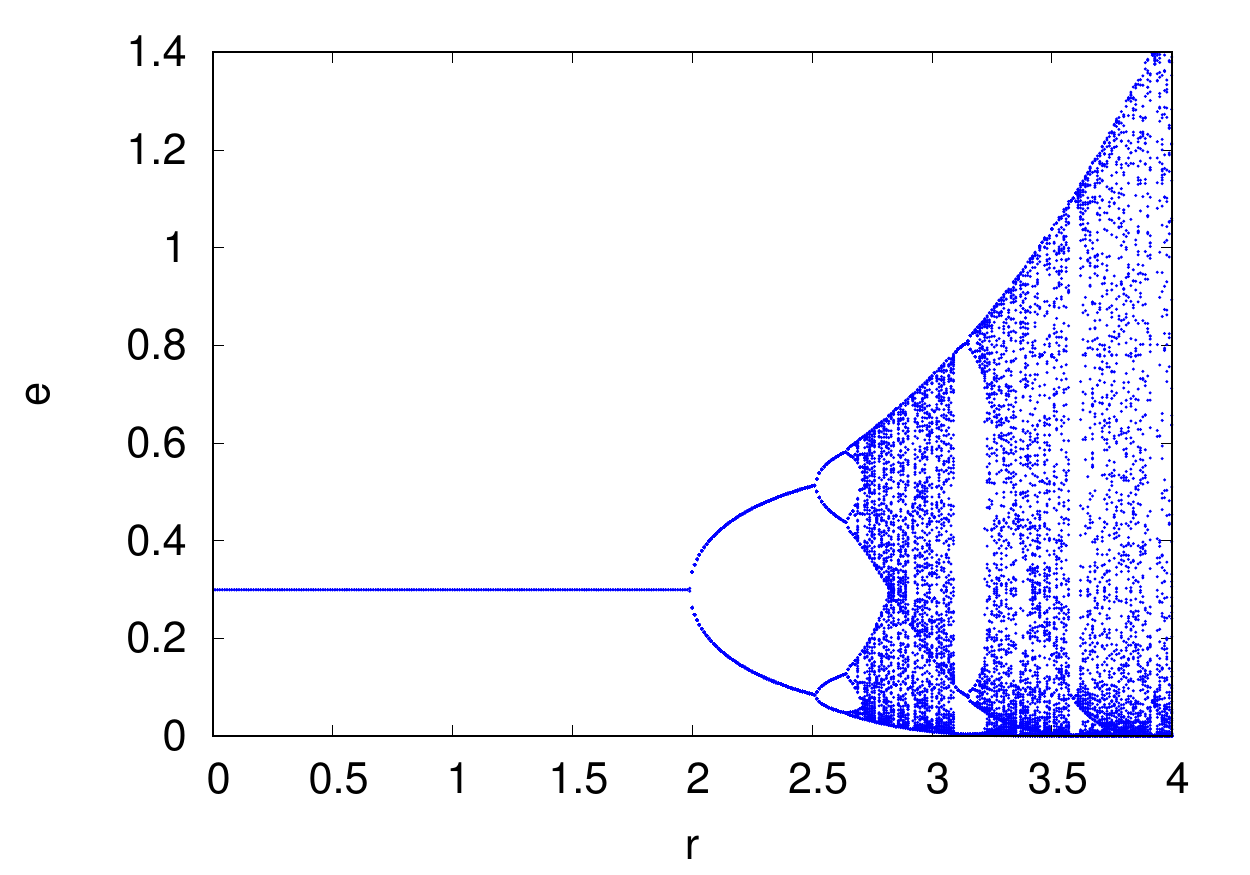}
\caption{Bifurcation diagram for the local population dynamics  (\ref{e})-(\ref{c}) with  parameters  
$\alpha = 0.1$,  $\delta= 0.01$ and $\rho = 0.005$. The points on the y-axis show 
the values of the engineer density  visited asymptotically from all initial conditions with $e_0 > 0$.}
\label{fig:1}
\end{figure}

The discrete-time population dynamics of the ecosystem engineers, given by the system of  recursion equations (\ref{e})-(\ref{c}), exhibits a complex dependence on the model parameters ($r, \delta, \rho$ and $ \alpha$) that was studied in great detail in \cite{Franco_17}. For instance, Figure \ref{fig:1} illustrates the  dependence  on the growth rate $r$ by showing the  bifurcation diagram for the engineer density  \cite{Sprott_03}.
 The  period-doubling bifurcation cascade is expected since the source of nonlinearity of the population dynamics  is Ricker's formula   \cite{Murray_03}.

Here our interest is on the nontrivial fixed-point solutions only, which are obtained by setting $e_{t+1} = e_t = e^*$, $h_{t+1} = h_t = h^*$ and $v_{t+1} = v_t = v^*$ in equations (\ref{e})-(\ref{c}). Assuming $e^* > 0$ we have  $h^* = e^*$  and 
$v^* = 1 - e^* \left ( 1 + \delta/\rho \right )$ with $e^*$ given by the solution of the  transcendental equation 

\begin{equation}\label{e*1}
\delta e^* = \left [ 1 - \exp \left ( - \alpha e^* \right ) \right ] \left [ 1 - e^* \left ( 1 + \delta/\rho  \right ) \right ] .
\end{equation}

\noindent  In the limit of small density, i.e.  $e^* \ll 1$,  this equation reduces to

\begin{equation}\label{e*2}
e^* \approx \frac{1 - \delta/\alpha}{1 + \delta/\rho},
\end{equation}

\noindent  and $v^* = \delta/\alpha$, indicating that this fixed point is physical for $\delta < \alpha $ only. As expected, $e^*$ increases with increasing $\alpha$ and $\rho$, and decreases with increasing $\delta$. Although eq.\ (\ref{e*1})  does not depend on the growth rate $r$,  large  values of this parameter lead to the instability of the  fixed point $e^*$ as illustrated in Figure \ref{fig:1}. Finally, the trivial fixed point $e^* = 0$, $h^* = 0$ and $v^* = 1$ is stable for $\delta > \alpha$. We refer the reader to Ref.\  \cite{Franco_17}
for the detailed analysis of the local stability of the fixed points of the recursion equations (\ref{e})-(\ref{c}) . 

\section{The coupled map lattice version of the Gurney and Lawton model }\label{sec:spatial}

The space-independent  recursion equations (\ref{e})-(\ref{c}) govern the local or single-patch population dynamics, where,  as already  pointed out, a patch  is  a  complete ecosystem (e.g., an  isle)  with the three types of habitats and a population of engineers. Those equations describe the growing phase of the population of engineers.  Here we introduce another phase - the  dispersal phase -  which we assume  takes place before the growing stage.  In particular,  we consider  a system of $N$  patches (e.g., an archipelago) and allow the engineers to circulate among neighboring patches, such that  a fraction $\mu$ of the population  in  patch $i$ is transferred to the $K_i$ neighboring patches. Hence after the dispersal stage the population at patch $i$ is  

\begin{equation}
E'_{i,t} = \left ( 1 - \mu \right ) E_{i,t} + \mu  \sum_j E_{j,t}/K_j,
\end{equation}

\noindent where  the sum is  over the $K_i$ nearest neighbors of patch $i$. For simplicity, we assume that the  total  number of habitats $T$  is the same for all patches, so that the effect of dispersal on the density of engineers is given by

\begin{equation}\label{el}
e'_{i,t} = \left ( 1 - \mu \right ) e_{i,t} + \mu  \sum_j e_{j,t}/K_j.
\end{equation}

\noindent for patch $i=1, \ldots, N$.  After dispersal of the engineers, the growing stage takes place within each patch according to the equations

\begin{eqnarray}
e_{i,t+1} &  =  & e'_{i,t} \exp \left [ r \left ( 1 - e'_{i,t}/h_{i,t} \right ) \right ]  \label{ei} \\
h_{i,t+1} & = & \left ( 1 - \delta \right ) h_{i,t} + c \left ( e'_{i,t} \right ) v_{i,t} \label{hi} \\
v_{i,t+1} & = &  \rho \left ( 1 - v_{i,t} -h_{i,t}  \right ) + \left [ 1- c \left ( e'_{i,t} \right ) \right ] v_{i,t}, \label{vi}
\end{eqnarray}

\noindent for $i=1, \ldots, N$. Together with eq.\ (\ref{el}), these equations  form  a coupled map lattice (see, e.g., \cite{Kaneko_92}) that describe the dynamics of the system of patches or metapopulation.

Since we are not interested on the formation of stationary spatial patterns, which appear only in the case the model parameters are such that the local dynamics (\ref{e})-(\ref{c}) is chaotic \cite{Franco_17}, the only patch arrangements we will consider in this paper are chains with an odd number of patches and reflective boundary conditions 
 (i.e., $K_1=K_N = 1$ and $K_i = 2, \forall i \neq 1,N$). In addition, we will focus on  a colonization or invasion scenario where at generation $t=0$ only the central patch $i_c= \left (  N + 1 \right ) /2$  of the chain is populated, whereas the other patches are composed entirely of virgin habitats (see, e.g.,  \cite{Comins_92}). In particular, we set $e_{i_c,0} = h_{i_c,0} = v_{i_c,0} = 0.5$ and $e_{i,0} = h_{i,0} =0$, $v_{i,0} =1$ for all $i \neq i_c$.

In the next section  we will study  the time dependence of the mean density of engineers,

\begin{equation}\label{emean}
\langle e_t \rangle =  \frac{1}{N} \sum_{i=1}^N e_{i,t}, 
\end{equation}

\noindent  and the mean fraction of virgin habitats,

\begin{equation}\label{emean}
\langle v_t \rangle =  \frac{1}{N} \sum_{i=1}^N v_{i,t}, 
\end{equation}

\noindent in the regime where the local dynamics is attracted to the nontrivial fixed point $e^* > 0$, so we 
 do not need to worry about accuracy issues caused by the chaotic amplification of numerical noise \cite{Rodrigues_15}.

\section{Results }\label{sec:res}

As  our focus is on the time dependence of  the global quantities $\langle e_t \rangle$  and $\langle v_t \rangle$, the 
results of this section are obtained solely through the numerical iteration of the coupled map lattice  equations (\ref{el})-(\ref{vi}). In addition, 
since we expect that  the time to reach the borders of the chain  scales linearly with the chain size $N$,  the  results  are presented in terms of the rescaled time $t/N$.

\begin{figure}[h]
\centering
\subfloat[{\bf (a)}]{\includegraphics[width=0.9\linewidth]{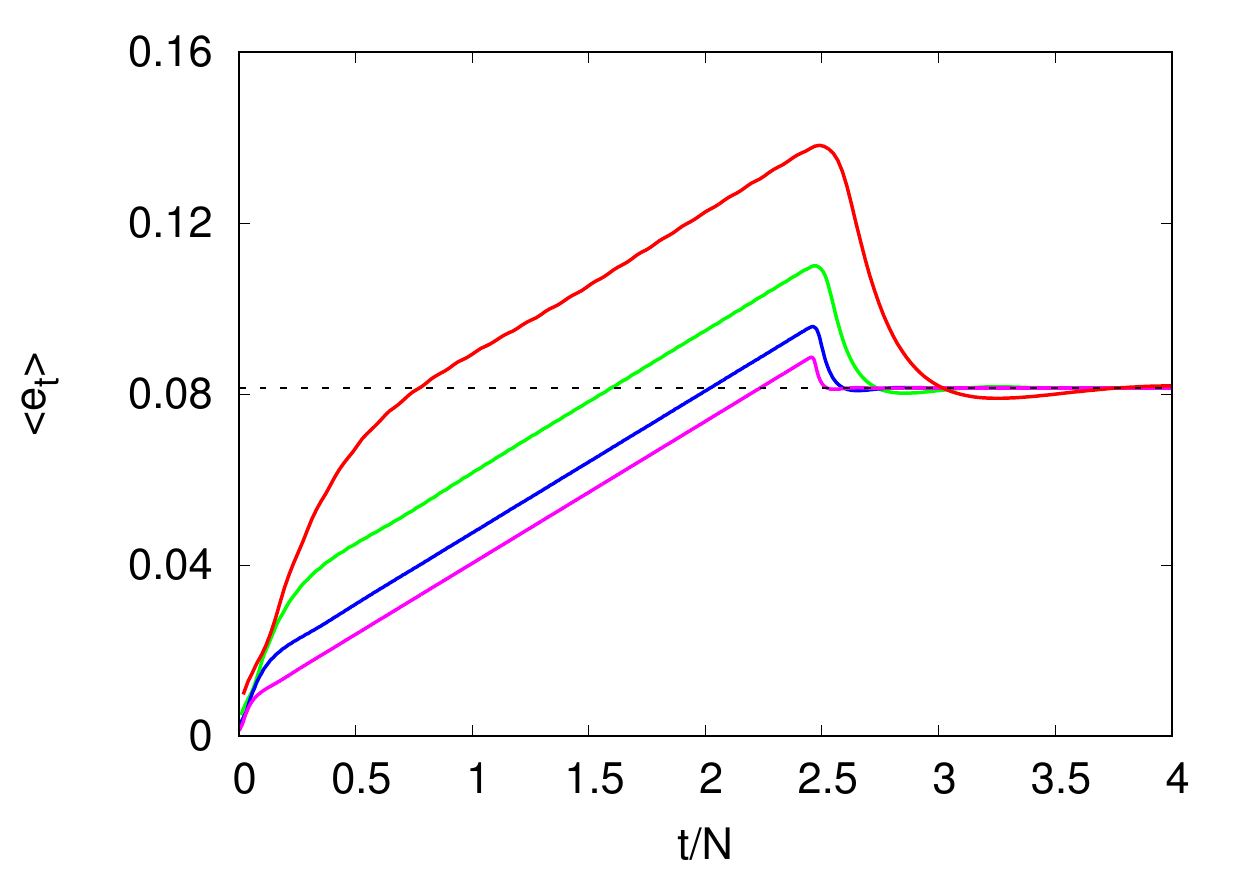}} \\
\subfloat[{\bf (b)}]{\includegraphics[width=0.9\linewidth]{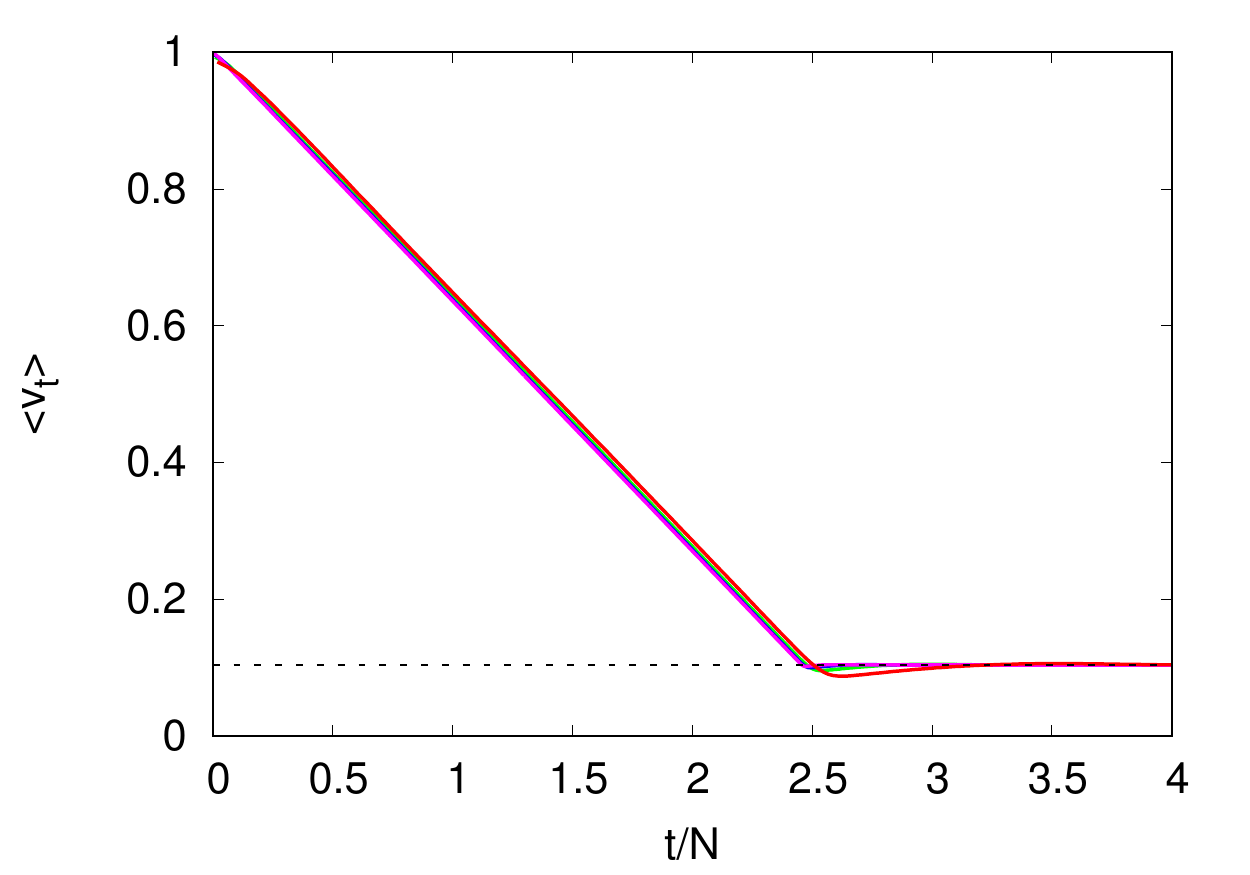}}
\caption{ (\textbf{a}) Mean engineer density $\langle e_t \rangle $ as function of the rescaled time $t/N$. 
(\textbf{b}) Mean fraction of virgin habitats $\langle v_t \rangle $ as function of the rescaled time $t/N$. The chain sizes are  $N=51$ (red line), $N=101$ (green line), $N=201$ (blue line) and $N=401$ (magenta line). The model   parameters  are
$r=1$, $\alpha = 1$,  $\delta= 0.1$, $\rho = 0.01$ and $\mu = 0.1$. The fixed point solution $e^* \approx 0.0814$ and $v^* \approx 0.1041$ obtained from eq. (\ref{e*1})  is shown by the dashed horizontal lines.}
\label{fig:2}
\end{figure}

\begin{figure}[h]
\centering
\includegraphics[width=0.9\linewidth]{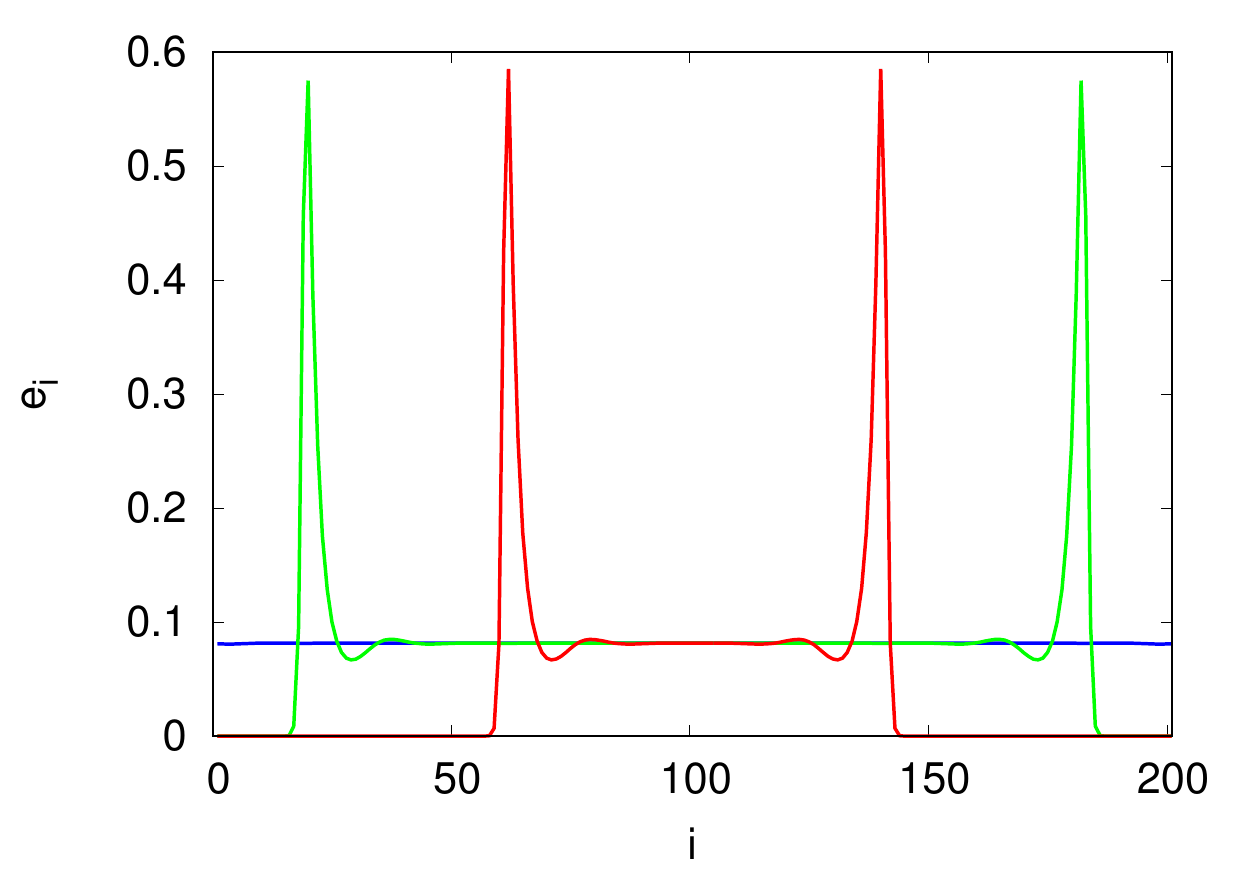}
\caption{Density of engineers in patch $i=1, \ldots, N$  at times $t/N= 1$ (red line), $t/N = 2$ (green line) and $t/N = 3$ (blue line). The chain size is $N=201$ and the model   parameters  are
$r=1$, $\alpha = 1$,  $\delta= 0.1$, $\rho = 0.01$ and $\mu = 0.1$.}
\label{fig:3}
\end{figure}

Figure \ref{fig:2}  shows the evolution of  the mean density of engineers and the mean fraction of virgin habitats for several  chain sizes $N$. It  reveals the dramatic effect of the engineers' mobility, which allow the population to reach densities well above those the environment could support in a situation of  equilibrium.   The initial increase of the mean density $\langle e_t \rangle$ reflects the expansion phase of the engineers,  which is accompanied by the monotone decreasing of the unexplored patches, as expected. This  expansion  halts only  when  the engineers  reach  the  borders of the chain   and the  end  of
the availability of unexplored virgin habitats results in a sharp drop on their density,  which then quickly converges to  the stationary value $\langle e_\infty \rangle = e^*$.   This scenario is corroborated by Figure \ref{fig:3} that shows the colonization wavefronts   at three distinct times.  The reason the size of the engineer density drop decreases with increasing $N$ (see panel (a) of Figure \ref{fig:2}) is simply  because the contribution of the two high-density wavefronts is watered down by the equilibrium-density of the bulk of the chain. We note that the shape and height of the wavefronts are not affected by the chain size. The time evolution of the mean fraction of usable habitat is qualitatively similar to that shown in panel (a)  for the density of engineers.

 \captionsetup[subfigure]{labelformat=empty}

Figure \ref{fig:2} suggests a direct way to calculate the mean speed $\nu $ of the colonization wavefronts, which is defined as  the ratio between the distance from the center to the borders of the chain (i.e., $N/2$) and the time $\hat{t}$ to reach those borders, i.e.,

\begin{equation}\label{speed}
\nu =  \frac{N}{2\hat{t}} .
\end{equation}

\noindent  In fact, $\hat{t}$ can be easily estimated by the time at which  $\langle e_t \rangle $ is maximum. For instance, for  all the chain sizes shown in Figure \ref{fig:2} we find  $\hat{t}/N \approx 2.46$, so that  $\nu \approx 0.20$. This means that  it is necessary about 5 generations  on the average   for the wavefront peak to move between contiguous patches. Since the mean speed of the colonization wavefronts is very weakly influenced by  the chain size, as illustrated in Figure \ref{fig:2}, henceforth we will consider chains of size $N=201$ only.
 
\begin{figure}[h]
\centering
\includegraphics[width=0.9\linewidth]{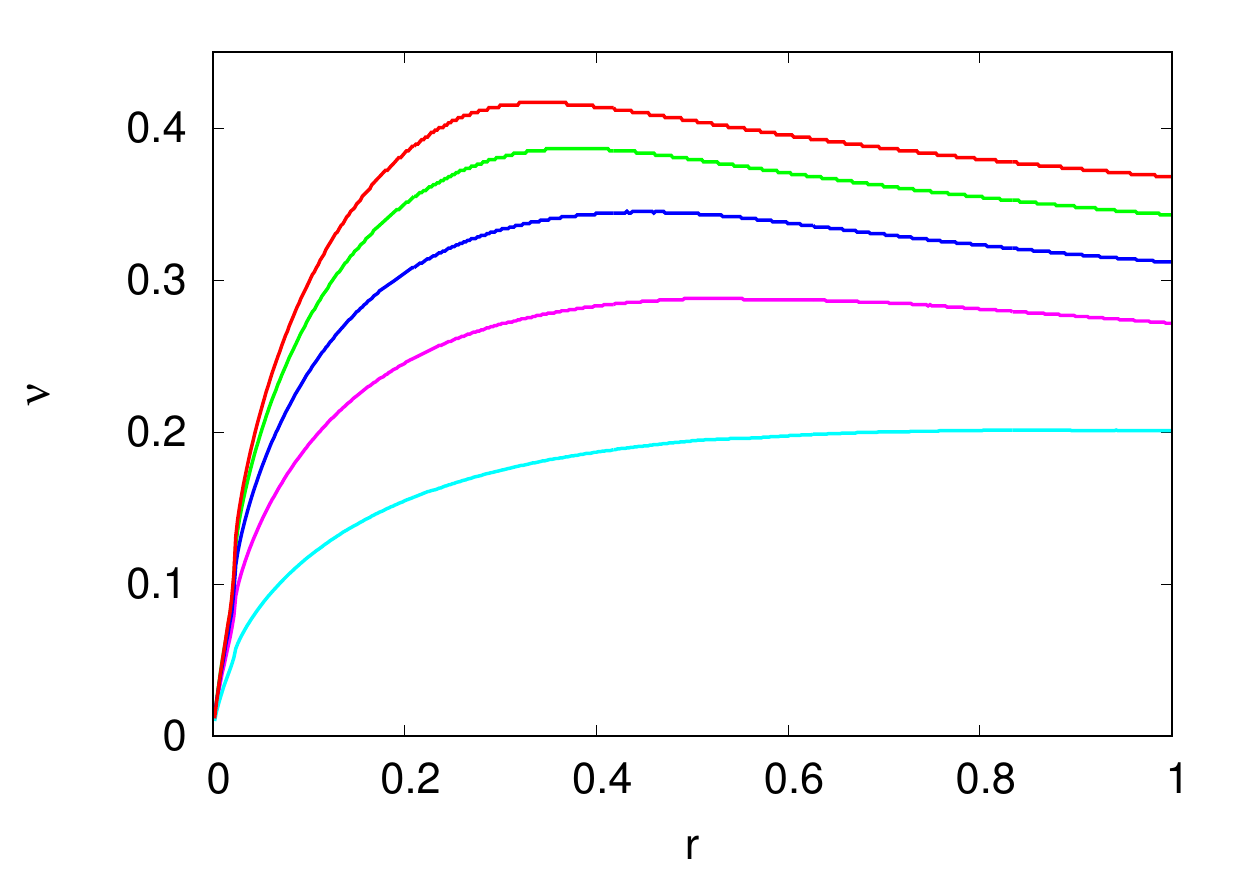}
\caption{Mean speed of the   colonization wavefronts $\nu$  as function of the growth rate $r$  for the dispersal probability
$\mu = 0.9$  (red line), $0.7$ (green line),  $0.5$ (blue line), $0.3$ (magenta line) and $0.1$ (cyan line). The chain size is $N=201$ and the model   parameters  are $\alpha = 1$,  $\delta= 0.1$, $\rho = 0.01$.}
\label{fig:4}
\end{figure}

 Figure \ref{fig:4} shows the dependence of the wavefront mean speed $\nu$ on the growth rate $r$ and on the dispersal probability $\mu$.
 Although these two parameters have no effect whatsoever on  the stationary solution $e^*$ and $v^*$, they have a strong  influence on the speed the population colonizes the unexplored patches. As expected, $\nu$ is a monotone increasing function of the dispersal probability $\mu$, but the rate of increase of $\nu$ decreases with increasing $\mu$, contrary to the unreflecting expectation of a linear relation between the speed of the wavefront and the dispersal probability. The dependence of $\nu$ on the growth rate $r$ is more interesting since it exhibits a non-monotone behavior, which is best seen in the figure  for large values of the dispersal probability  but that actually happens for all values of $\mu$. Since only a fast growing population can guarantee  a very large density at the borders of the expanding colony (see Figure \ref{fig:3}) and hence take advantage of the  neighboring unexplored  patches, one should expect a steep increase of  $\nu$  with increasing $r$, despite the fact that  $r$ plays no role in the equilibrium situation.   In fact, this is what one observes in  Figure \ref{fig:4},  provided that $r$ is not too large.   The smooth decrease of $\nu$ for large $r$ is probably due to the negative feedback of a  large growth rate on the engineer population when the available fraction of useful habitats is not large enough. 
 In fact, the maximum wavefront speed observed in the figure is a result of a fine tuning between the growth rate $r$ and the potential to create usable habitats $\alpha$ from the virgin patches.
Interestingly, although the first engineers to reach the virgin patches are doomed to extinction because there are no usable habitats in those patches (see eq.\ (\ref{ei})), they build the usable habitats (see eq.\ (\ref{vi})) for the next wave of migrants.
   
\begin{figure}[h]
\centering
\includegraphics[width=0.9\linewidth]{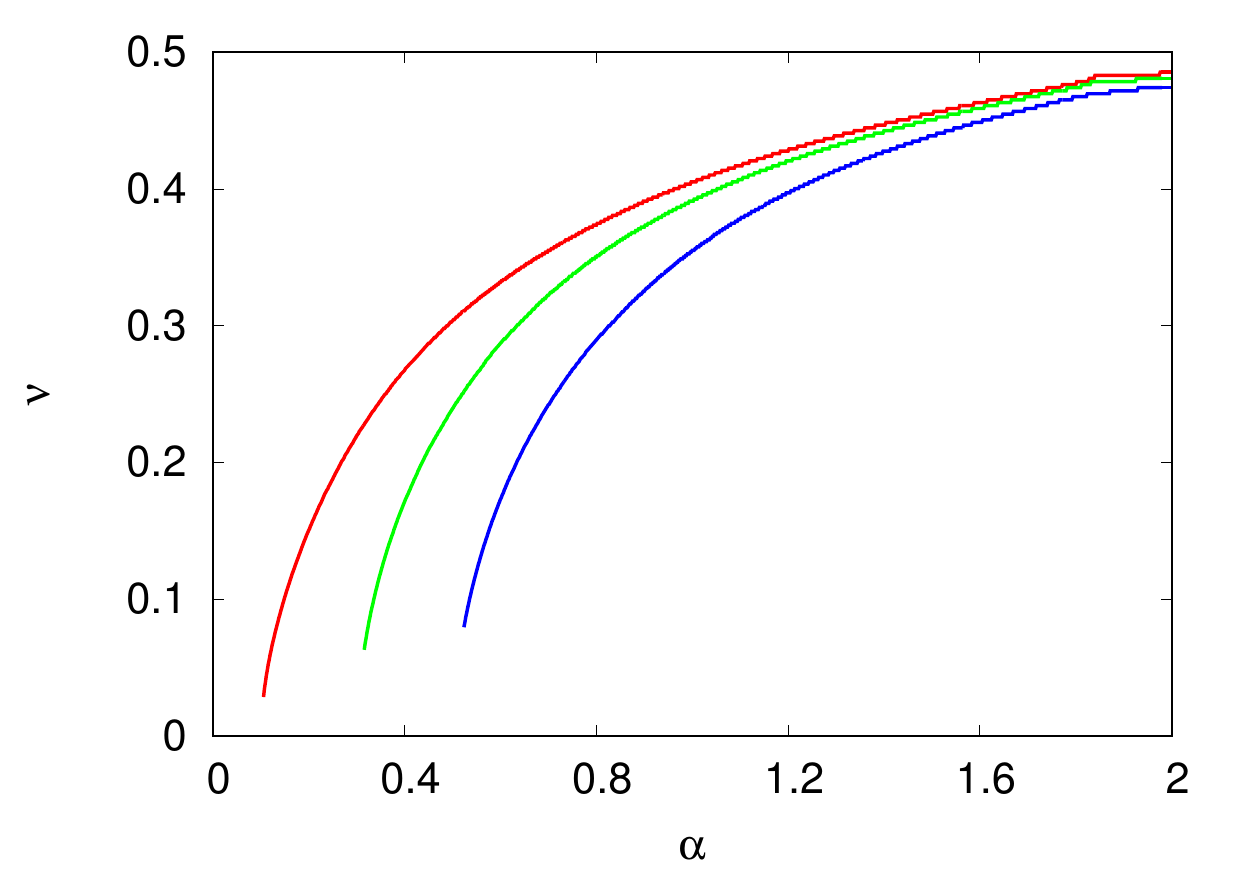}
\caption{Mean speed of the   colonization wavefronts  $\nu$ as function of the productivity rate $\alpha$  for the decay probability 
$\delta = 0.1$  (red line), $0.3$ (green line), and   $0.5$ (blue line). The curves are shown for $\alpha > \delta$ so that the population is viable at  equilibrium, i.e., $e^* > 0$. The chain size is $N=201$ and the model   parameters  are $r=0.5$, $\alpha = 1$, $\rho = 0.01$ and  $\mu = 0.9$.}
\label{fig:5}
\end{figure}

We have verified that the mean speed $\nu$ of the colonization wavefronts is not influenced by the recovery probability $\rho$ of the degraded habitats, as expected. In fact, the capacity of recovery of the degraded habitats that are left far behind the invasion front is completely  irrelevant for the survival and growth of the pioneers in the colonization front.  The dependence of $\nu$ on the productivity rate $\alpha$ and on the decay probability $\delta$ is summarized in Figure \ref{fig:5}. The results are restricted to the region $\alpha > \delta$ where we can guarantee the existence of a viable equilibrium population of engineers, i.e.,  $e^* > 0$. The mean speed $\nu$ increases monotonically 
with $\alpha$ since the efficiency of the transformation of virgin habitats into usable habitats is crucial for the survival  of the second wave of migrants in the colonization front, as pointed out before. In the same line of reasoning, if the recently created usable habitats decay too rapidly, then the  colonization front will be delayed as shown in  Figure \ref{fig:5}. However, if the productivity $\alpha$ is large then the decay probability $\delta$ has only a negligible retarding effect on the  mean speed of the wavefront.

\captionsetup[subfigure]{labelformat=empty}
\begin{figure}[h]
\centering
\subfloat[{\bf (a)}]{\includegraphics[width=0.9\linewidth]{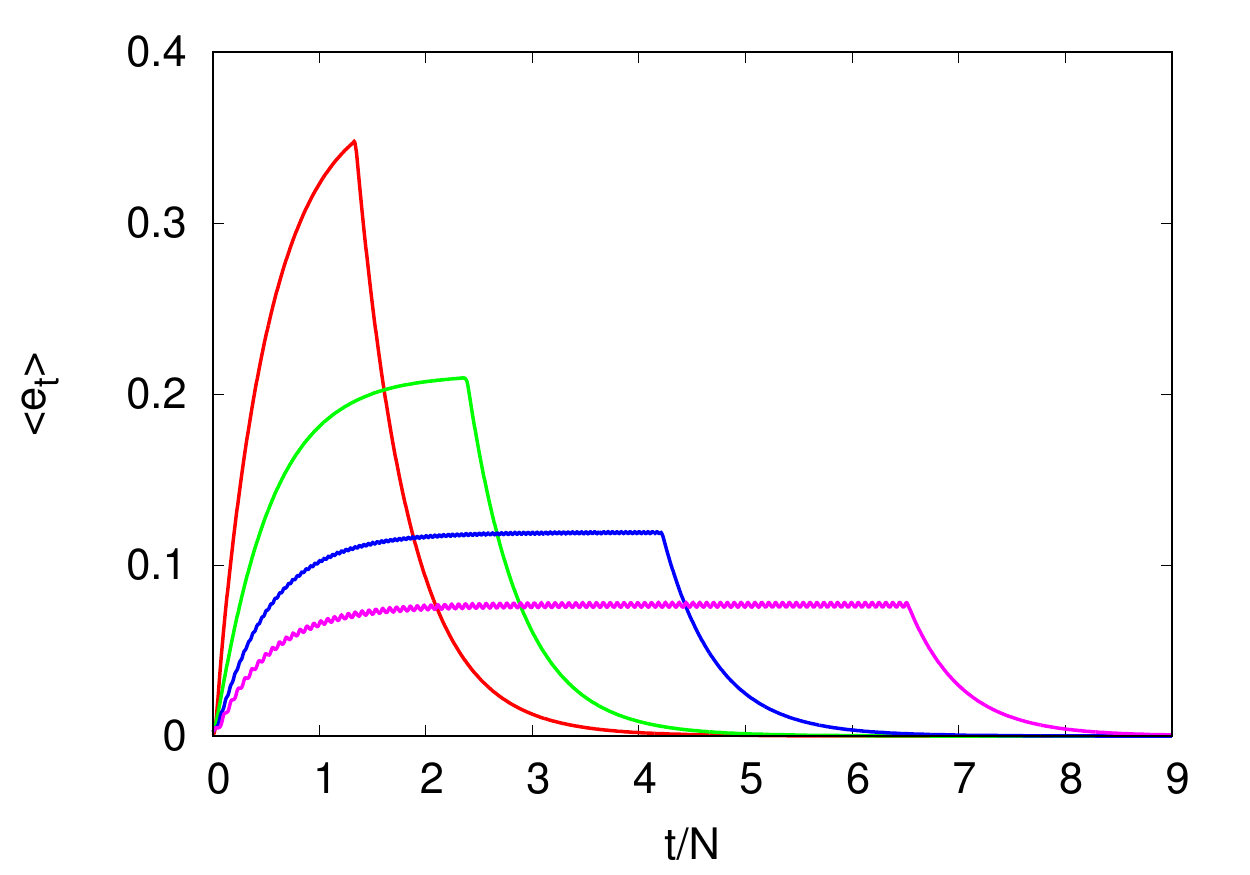}}\\
\subfloat[{\bf (b)}]{\includegraphics[width=0.9\linewidth]{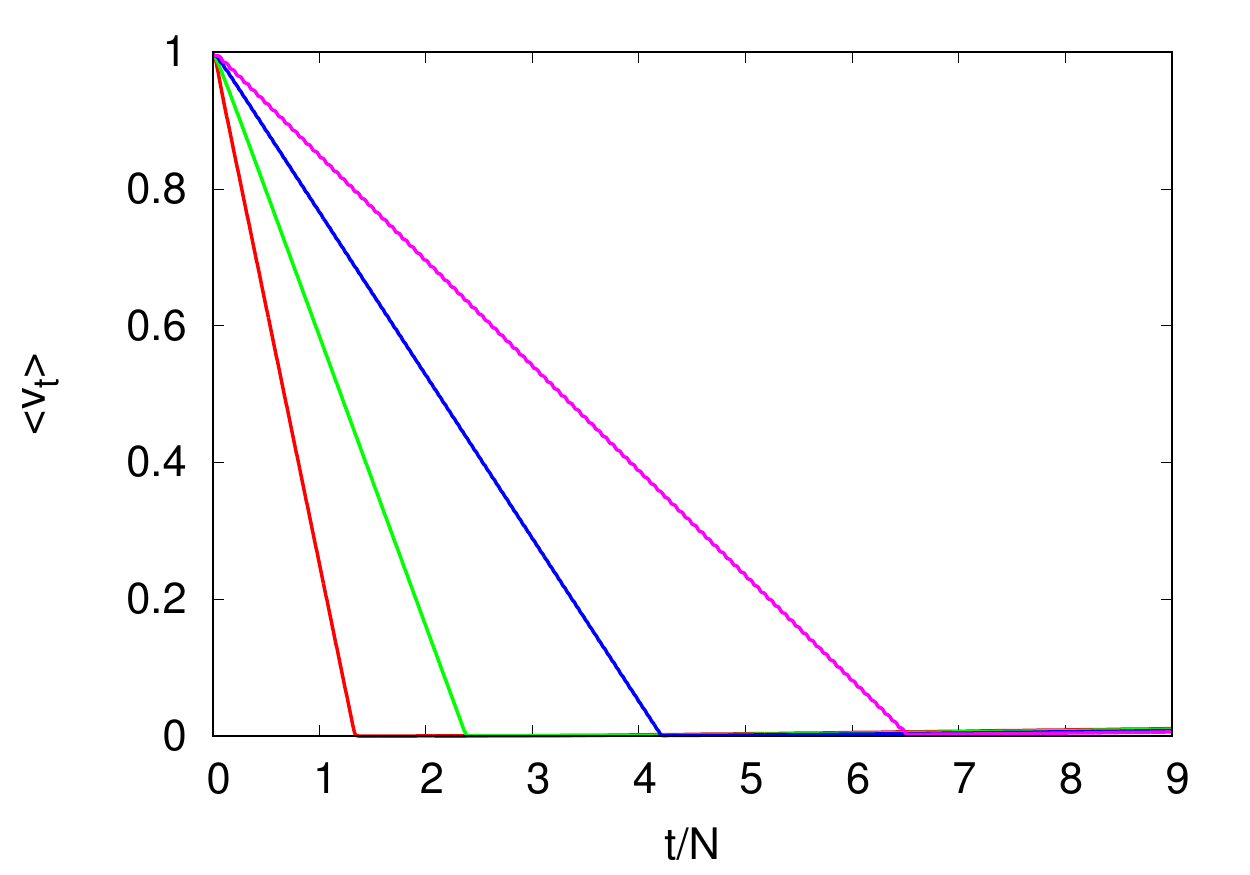}}
\caption{ (\textbf{a}) Mean engineer density $\langle e_t \rangle $ as function of the rescaled time $t/N$. 
(\textbf{b}) Mean fraction of virgin habitats $\langle v_t \rangle $ as function of the rescaled time $t/N$. The dispersal probability is
 $\mu = 0.9$ (red line), $\mu = 0.1$ (green line), $\mu = 0.01$ (blue line) and $\mu = 0.001$ (magenta line) and the other model   parameters  are $r=1$, $\alpha = 1$,  $\delta= 0.01$, $\rho = 10^{-5}$.  The chain size is $N=201$ and the fixed point solution is $e^* \approx 0.001$ and $v^* \approx 0.01$.}
\label{fig:6}
\end{figure}   

The collapse of the population, which happens when there are no more  virgin patches to be explored, can be made more spectacular by setting the model parameters such that the equilibrium population density is very small, i.e., $e^* \ll 1$. One way to achieve this is  by setting  $\rho \ll 1$ (see eq.\ (\ref{e*2})),  resulting in the global measures  $\langle e_t \rangle $ and $\langle v_t \rangle $  displayed in Figure \ref{fig:6}. 
Let us consider first panel (b) of Figure \ref{fig:6} that shows  the linear decrease of the fraction of virgin habitats   with  time,  which ends when the colonization fronts reach the borders of the chain at time $t = \hat{t}$. At this moment we have  $\langle v_{\hat{t}} \rangle \ll v^* $ and from then on the fraction of virgin habitats begins to increase very slowly following the time scale set by the recovery probability $\rho$ (this last stage is barely seen on the scale of the figure).

The time dependence of the density of engineers shown in panel (a) of Figure \ref{fig:6}  is more instructive. The population with the highest mobility uses up the environmental  resources quickly and   reaches very  high densities before plunging towards the  low equilibrium density, whereas the population with the lowest mobility can maintain an average density value for a long time  before exhausting the resources. It is interesting    that in this case the mean density of engineers  exhibits a sort of metastable equilibrium (i.e., $\langle e_t \rangle $ is practically constant for a long period of time) although the population is expanding  through the unexplored patches. 

\begin{figure}[h]
\centering
\includegraphics[width=0.9\linewidth]{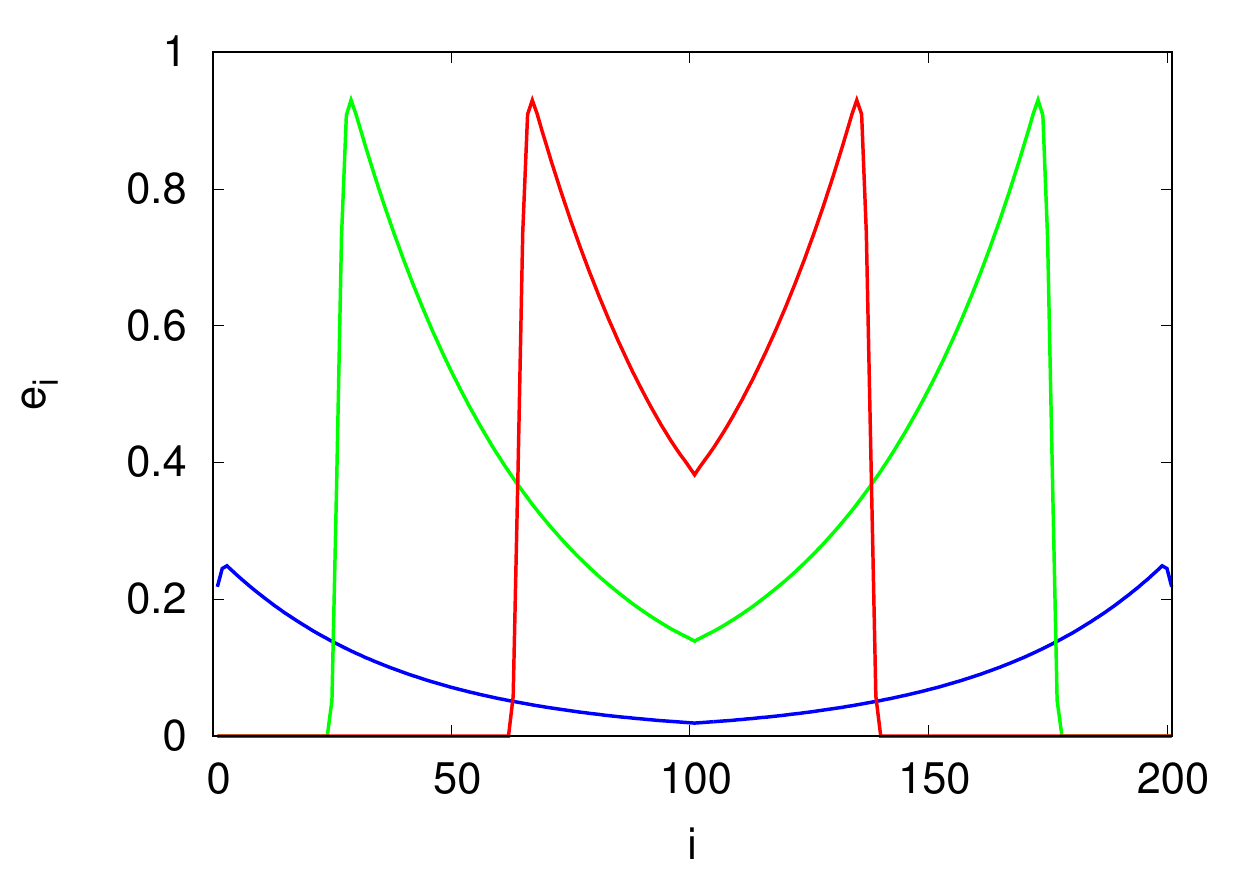}
\caption{Density of engineers in patch $i=1, \ldots, N$  at times $t/N= 0.5$ (red line), $t/N = 1$ (green line) and $t/N = 2$ (blue line). The chain size is $N=201$ and the model   parameters  are
$r=1$, $\alpha = 1$,  $\delta= 0.01$, $\rho = 10^{-5}$ and $\mu = 0.9$.}
\label{fig:7}
\end{figure}   

\begin{figure}[h]
\centering
\includegraphics[width=0.9\linewidth]{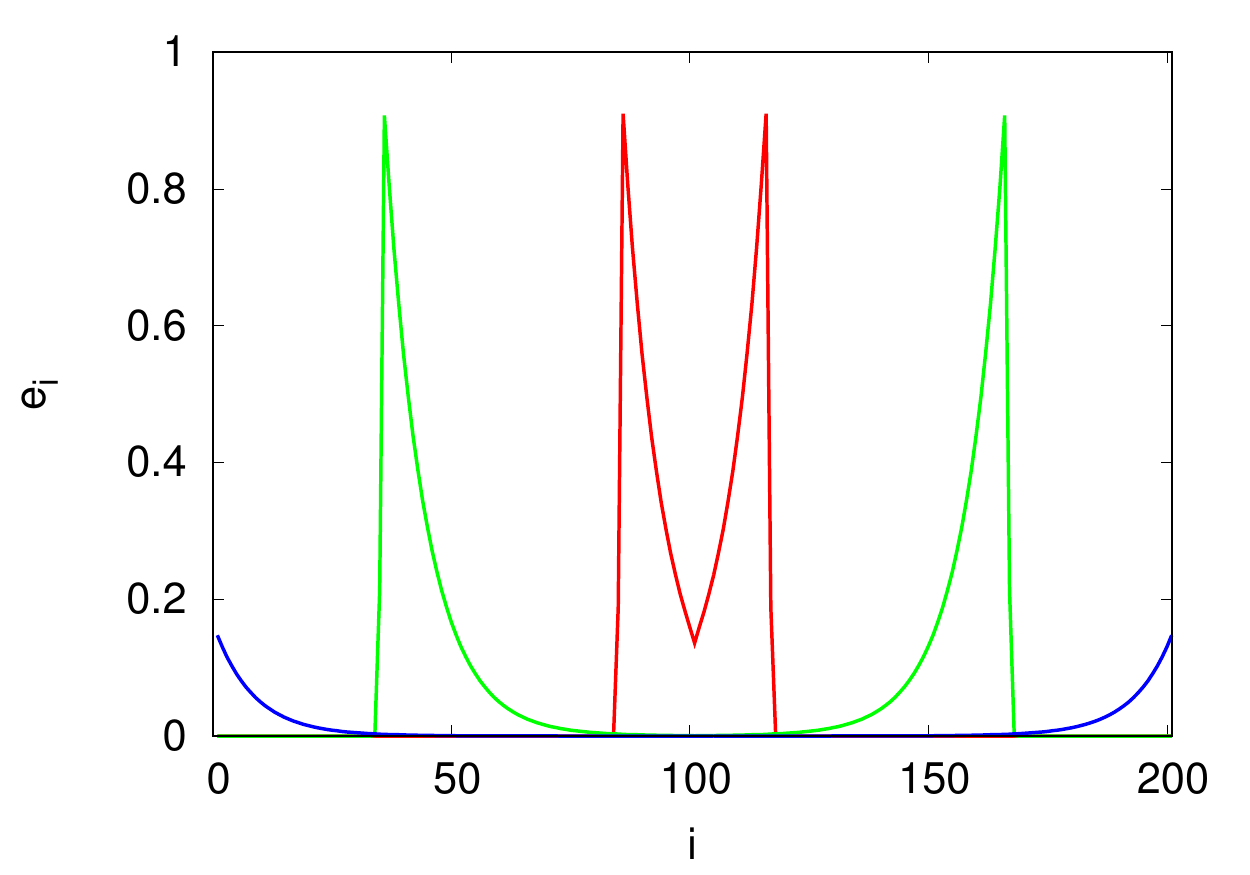}
\caption{Density of engineers in patch $i=1, \ldots, N$  at times $t/N= 1$ (red line), $t/N = 4$ (green line) and $t/N = 7$ (blue line). The chain size is $N=201$ and the model   parameters  are
$r=1$, $\alpha = 1$,  $\delta= 0.01$, $\rho = 10^{-5}$ and $\mu = 0.001$.}
\label{fig:8}
\end{figure}   

The qualitative differences of the 
population density in these mobility extremes  are  easily understood with the aid of Figures \ref{fig:7} and \ref{fig:8} that show the population densities in each patch for $\mu = 0.9$ and $\mu = 0.001$, respectively. In fact, the reason the high mobility population (see Figure \ref{fig:7})  attains such high densities  before it collapses is simply that it reaches the borders  before the usable habitats in the center of the chain can degrade appreciably. In the low mobility case (see Figure \ref{fig:8}), the center of the chain is practically desert when the colonization front is moving towards the borders.  The metastable equilibrium mentioned before is simply a consequence of the invariance of the shape of the  wavefronts, as the colonization fronts are the only places where usable habitats are found. (The spatial distribution of usable habitats is
indistinguishable from that of the engineers.) 
  We note that  the density is higher near the borders than near the center of the chain because the usable habitats
at the borders were created much later than those in the center and so had  less time to decay into the degraded class.

 Finally, we note that in the case of unviable patches, i.e., in the  regime  $\alpha < \delta$ where the nontrivial fixed point $e^* > 0$ is unphysical and 
$e^*=0$ is stable, the colonization of the unexplored virgin patches fails and the engineers quickly go   extinct. The reason is 
that the  vast supply of  unexplored patches  is irrelevant  if the  production of usable habitats from them  is not enough to balance the decay of the usable habitats  into degraded habitats.

\section{Discussion}\label{sec:disc}

The increase of the  mean density of engineers $\langle e_t \rangle $  during the  expansion phase of the colony is totally expected, of course, since as new patches are invaded by the engineers their overall density must increase. (We recall that in our scenario $\langle e_t \rangle $ varies from the arbitrarily set initial value $\langle e_0 \rangle = 0.5/N $ to  the local equilibrium density $\langle e_\infty \rangle = e^*$, where $N$ is the number of patches.)  What is  surprising is that the transient density reaches values much larger than the equilibrium 
density and plunges sharply when the colonization front hits the  boundary of the available space (see, e.g.,  Figure \ref{fig:6}). This phenomenon was observed en passant in Ref.\ \cite{Franco_17} for the regime of  chaotic local dynamics, and was somewhat concealed by the difficulty of  controlling the numerical accuracy of hundreds of chaotically oscillating coupled patches. Here we have offered a detailed analysis of the colonization process  for the numerically more manageable  dynamic regime  where  the sole attractors  are fixed points.

We note that if we set the recovery probability $\rho$ of the degraded habitats to a small value, then the equilibrium population density $e^* \approx \rho \left ( 1/\delta - 1/\alpha \right )$ will be small too (see eq.\ (\ref{e*2})), and since the maximum mean density $\langle e_{\hat{t}} \rangle$  does not depend on  $\rho$,  we get $e^*/\langle e_{\hat{t}} \rangle \sim \rho $, which implies  a catastrophic reduction of the population density around time $\hat{t}$. This is the reason we liken 
the sharp drop of the population density when the supply of unexplored resources is exhausted (see Figures \ref{fig:2} and \ref{fig:6}) to the collapse of ancient human societies  that   overexploited their environment \cite{Diamond_05}. 

In the context of the collapse of human societies, our  population dynamics model produces two outcomes that are worth emphasizing.  The first result is that   the disaster comes   without warning  since the shape and the height of the wavefront 
 are constant before the colonization front hits the chain border  (see the curves for the low mobility values of Figures  \ref{fig:6} and \ref{fig:8}), whereas  the  mean density $ \langle e_{t} \rangle$  is either increasing or practically constant  before the collapse (see Figures \ref{fig:2} and  \ref{fig:6}).
This accords with  Diamond's interpretation of the archeological records of collapsed civilizations  \cite{Diamond_05}: ``In fact, one of the main lessons to be learned from the collapses of the Maya, Anasazi, Easter Islanders, and those other past societies (as well as from the recent collapse of the Soviet Union) is that a society's steep decline may begin only a decade or two after the society reaches its peak numbers, wealth, and power.''

The second  consequence of our model is that  overpopulation is a natural outcome of the nonlinear dynamics of the ecosystem engineer population expanding over unexplored habitats. A rough  global measure of the overpopulation  at  time $t$  is given by the ratio $\langle e_t \rangle/e^*$ that equals 1 in the equilibrium situation. We note, however,  that the  local engineer density in the patches that are part of the colonization front are much higher than the overall mean density (see Figures \ref{fig:3}, \ref{fig:7} and  \ref{fig:8}). This is so because the second wave of migrants finds empty patches composed  mostly of usable  habitats (meaning a large carrying capacity) that resulted from the  
work of the extinct first  wave of migrants  on the original virgin habitats.  

We find it quite remarkable that the model proposed by Gurney and  Lawton to study the  population dynamics of ecosystem engineers \cite{Gurney_96},  which seems to have been developed with an eye on the ecology of beavers  \cite{Wright_04}, could provide  such interesting insights on the collapse dynamics of past human societies, without incorporating  specific traits of those societies \cite{Diamond_05}.  For instance, one such a trait is   existence of ruling elites that parasitize on the large mass of producers (commoners), using their workforce to produce luxury items and  religious monuments \cite{Motesharrei_14}. This feature could easily be incorporated in our model by requiring that only the commoners  modify the virgin habitats and that the elite members use a disproportionally large amount  of usable habitats. Nonetheless, our results show that the collapse of an expanding population of ecosystem engineers seems to be a robust, unavoidable consequence of the nonlinear feedback between the population and environment variables, so a more detailed modeling of  human societies will probably  have little effect on our findings.

\acknowledgments
This research was  supported in part by grant
15/21689-2, Funda\c{c}\~ao de Amparo \`a Pesquisa do Estado de S\~ao Paulo 
(FAPESP) and by grant 303979/2013-5, Conselho Nacional de Desenvolvimento 
Cient\'{\i}\-fi\-co e Tecnol\'ogico (CNPq).

\end{document}